\documentclass{sigchi}

\toappear{Permission to make digital or hard copies of all or part of this work for personal or classroom use is granted without fee provided that copies are not made or distributed for profit or commercial advantage and that copies bear this notice and the full citation on the first page. Copyrights for components of this work owned by others than ACM must be honored. Abstracting with credit is permitted. To copy otherwise, or republish, to post on servers or to redistribute to lists, requires prior specific permission and/or a fee. Request permissions from permissions@acm.org. \\
{\emph{CHI'17}}, May 6--11, 2017, Denver, CO, USA \\
Copyright \copyright~2017 ACM. ISBN 978-1-4503-4655-9/17/0\$15.00 \\
DOI: http://dx.doi.org/10.1145/3025453.3025930}

\usepackage{balance} 
\usepackage{graphics} 
\usepackage[T1]{fontenc}
\usepackage{txfonts}
\usepackage{mathptmx}
\usepackage[pdftex]{hyperref}
\usepackage{color}
\usepackage{booktabs}
\usepackage{textcomp}
\usepackage{microtype} 
\usepackage{ccicons} 
\usepackage{color}

\usepackage{todonotes}

\def\plaintitle{Scalable Annotation of Fine-Grained Categories Without Experts}
\def\plainauthor{Timnit Gebru$^1$, Jonathan Krause$^1$, Jia Deng$^2$, Li Fei-Fei$^1$\\
\affaddr{Stanford University$^1$, University of Michigan$^2$}\\
\affaddr{$\{$tgebru, jkrause, feifeili$\}$@cs.stanford.edu, jiadeng@umich.edu}
}

\def\plainkeywords{Fine-grained dataset; crowdsourcing; human computation}

\makeatletter
\def\url@leostyle{%
 \@ifundefined{selectfont}{
 \def\UrlFont{\sf}
 }{
 \def\UrlFont{\small\bf\ttfamily}
 }}
\makeatother
\def\pprw{8.5in}
\def\pprh{11in}

\definecolor{linkColor}{RGB}{6,125,233}
\hypersetup{%
 pdftitle={\plaintitle},
pdfauthor={\plainauthor},
 pdfkeywords={\plainkeywords},
 bookmarksnumbered,
 pdfstartview={FitH},
 colorlinks,
 citecolor=black,
 filecolor=black,
 linkcolor=black,
 urlcolor=linkColor,
 breaklinks=true,
}

\begin{document}

\title{\plaintitle}
\author{\plainauthor}

\maketitle

\begin{abstract}
We present a crowdsourcing workflow to collect image annotations for visually similar synthetic categories without requiring experts. In animals, there is a direct link between taxonomy and visual similarity: e.g. a collie (type of dog) looks more similar to other collies (e.g. smooth collie) than a greyhound (another type of dog). However, in synthetic categories such as cars, objects with similar taxonomy can have very different appearance: e.g. a 2011 Ford F-150 Supercrew-HD looks the same as a 2011 Ford F-150 Supercrew-LL but very different from a 2011 Ford F-150 Supercrew-SVT. We introduce a graph based crowdsourcing algorithm to automatically group visually indistinguishable objects together. Using our workflow, we label $712,430$ images by $\sim1,000$ Amazon Mechanical Turk workers; resulting in the largest fine-grained visual dataset reported to date with $2,657$ categories of cars annotated at $1/20^{th}$ the cost of hiring experts.

\end{abstract}
\category{Algorithm}{Information Interfaces and Presentation
 (e.g. HCI)}{Miscellaneous} 


\section{Introduction}
To automatically detect cancer, an image recognition system has to be able to distinguish between cancerous and noncancerous cells which look almost identical. This system needs to perform what is called fine-grained image classification: distinguishing between highly similar objects~\cite{fg1,fg2,birdsnap,fg4,fg5}. Recently, deep learning based classification models have been shown to distinguish between similar categories of objects (e.g. birds) with greater than 92\% accuracy if trained with enough labeled data~\cite{jon}. Thus, the first step in creating any such image classification system is gathering sufficient annotated data where each image is labeled with the correct category.

\begin{figure}
\centering
 \includegraphics[width=0.9\columnwidth]{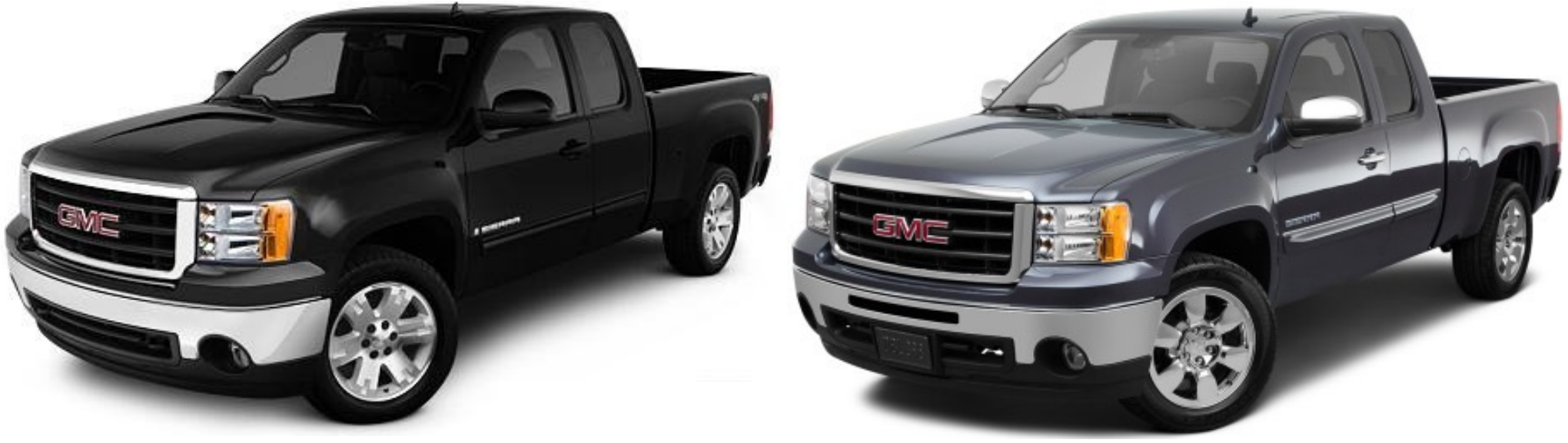}
 \caption{Fine-grained car categories look very similar to a non-expert. Here, a 2011 GMC Sierra-1500 extended cab sle (left) and a 2011 GMC Sierra-1500 extended cab work truck (right) look almost identical. However, the second car has a small opening on the metal bumper whereas the first one does not. Our fine-grained object grouping methodology correctly assigns these cars to different categories. }~\label{fig:figure1}
\end{figure}

Existing large scale visual datasets, such as ImageNet, were constructed using crowdsourcing to label the object instances in images~\cite{imagenet}. These objects range from animals like cats or dogs to synthetic artifacts like airplanes and cars. A worker annotating images for this dataset only needs to be able to distinguish between very dissimilar categories, like cats and dogs or cars and airplanes. This allows any non-expert worker to label images in ImageNet with no prior training. On the other hand, a non-expert cannot accurately distinguish between fine-grained categories like cancerous and non-cancerous cells or 2011 GMC Sierra-1500 extended cab sle and 2011 GMC Sierra-1500 extended cab work truck (Figure~\ref{fig:figure1}). Thus, building fine-grained datasets is prohibitively expensive because hiring experts to annotate images costs a conservative average of $16$ cents per annotation (assuming a wage of \$10 per hour at an annotation speed of 1 image per minute). For instance, there are over $2,500$ visually distinct types of cars manufactured since 1990 alone~\cite{edmunds}, and typical large scale image classification systems require over $1,000$ annotated training instances per category~\cite{vgg}. To build a comprehensive fine-grained car classification system, we would need to collect a dataset with over 2 million annotations, a process which would cost over $\$300,000$ using experts. Thus, inexpensively collecting large fine-grained datasets requires a methodology that is able to utilize non-expert workers.

In this paper, we present a crowdsourcing workflow to collect visual datasets of synthetic fine-grained categories without requiring experts. The first and most difficult step in our workflow consists of creating a class list, i.e. grouping visually indistinguishable objects into the same category. For a dataset of all cars manufactured since 1990, this step consists of examining $15,213$ different types of cars and grouping those that have the same appearance together. This step is especially difficult for fine-grained datasets where different categories look very similar. In our case, care must be taken to ensure that cars with slight visual distinctions (such as those in Figure~\ref{fig:figure1}) are not grouped into the same class. We introduce a graph based crowdsourcing algorithm to cluster visually indistinguishable categories of cars. First, we construct a tree using existing car taxonomy where the leaves correspond to different categories. We then use binary tasks to iteratively merge pairs of leaves moving up the tree in a bottom-up-breadth-first search order. After iteratively merging car categories, we use connected component analysis~\cite{connected-components} to arrive at our final groupings of cars.
\begin{figure}
 \centering
 \includegraphics[width=0.9\columnwidth]{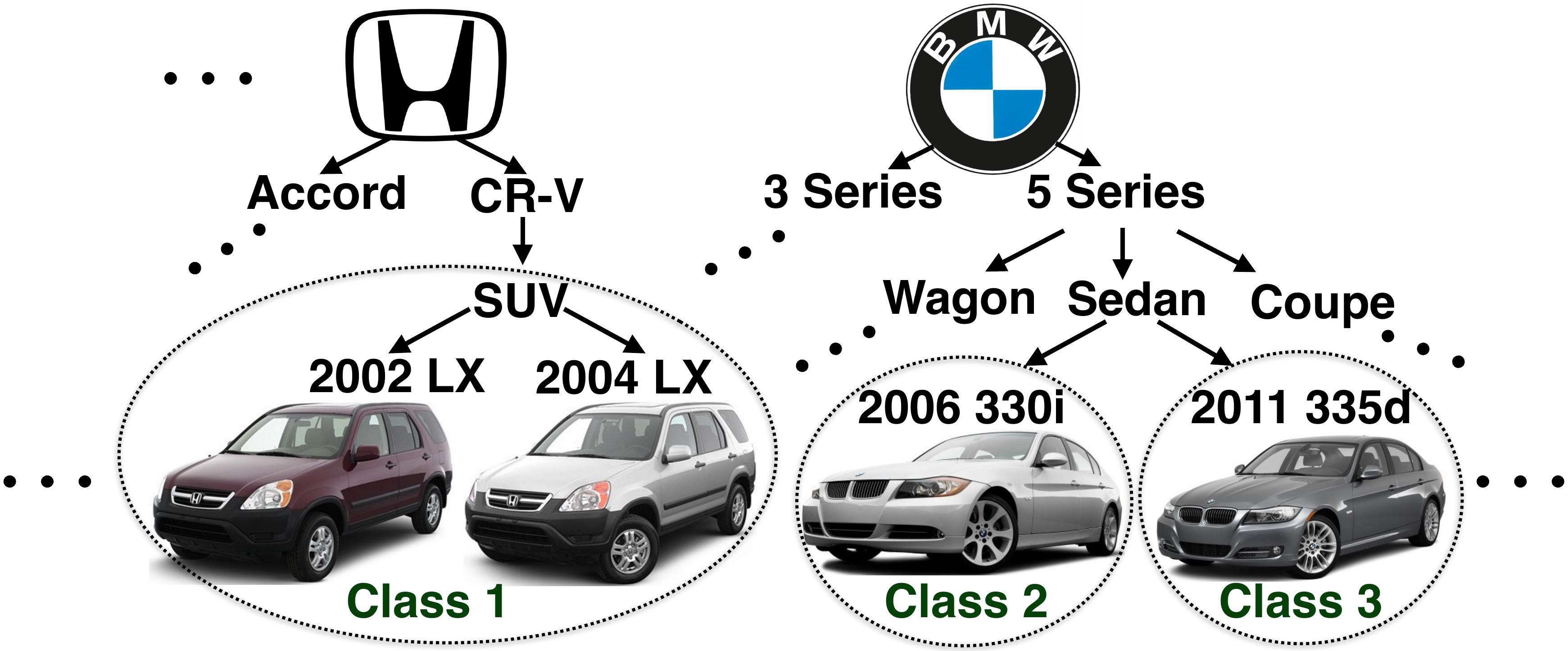}
 \caption{Cars follow a taxonomic hierarchy which can be interpreted as a tree. At the root node is the make of the car, e.g. Honda vs BMW, followed by the model, body type (e.g. sedan, wagon, coupe, SUV), year and trim (e.g. LX, 330i). All nodes under one make tree (e.g. BMW) are visually different from all nodes under another make tree (e.g. Honda). But nodes under the same tree can be visually the same or different. In this example, the two cars at the BMW leaf node have slight visual differences and are grouped into two different classes. Conversely, those under the Honda leaf node are grouped into the same class.}~\label{fig:car_hierarchy}
\end{figure}

Finally, we use our generated class list to collect images for all the car categories by querying craigslist.com and cars.com -- websites where people sell their cars along with images and detailed description of the car's make and model. For each type of car, we use text processing methods to parse cragislist.com and cars.com posts and find vehicles that match our query. We then download all images for that post and use an Amazon Mechanical Turk (AMT) verification task to filter out images that do not have cars. Each image contains one car and is annotated by its corresponding category from our constructed class list. Our result is a fine-grained car dataset with $2,657$ categories and $712,430$ images with a precision of $96.6\%$. It was labeled by over a thousand AMT workers and cost only $\$6,000$ instead of an estimated $\sim\$119,000$ had we hired experts. We evaluate our visual taxonomy creation algorithm on a subset of the dataset consisting of all Honda Accord sedans manufactured since 1990, and show comparable accuracy with an expert.

\section{Related Work}
A number of fine-grained datasets have been released for image classification. The most widely used one, CUB-200, consists of $200$ species of birds with $30$ images per category~\cite{cubs}. Recently, larger datasets of birds have been constructed by crowdsourcing the task to experts~\cite{visipedia,birdsnap}. To date, the largest fine-grained dataset of synthetic objects is CompCars consisting of $214,345$ images of $1,687$ car models~\cite{compcars}. However, its use in fine-grained image classification is limited because all cars for one model are grouped into the same category, in spite of their visual differences. For instance, a 2005 Toyota Corolla looks very different from a 2015 Toyota Corolla but is categorized into the same class in the CompCars dataset.

Our proposed workflow draws inspiration from prior work in decreasing image labeling cost through crowdsourcing while maintaining high annotation precision. Previous works have explored methods to speed up binary annotation tasks by minimizing penalties for worker errors~\cite{ranjay}. Others have exploited relationships between categories for efficient multi-label annotations~\cite{jia}. Methods to acquire expert level labels through crowdsourcing have also been extensively studied by investigating the use of multiple noisy workers to annotate data~\cite{welinder,sheng}. In all these works, crowdsourcing has proved to be a reliable technique to obtain labels with expert level accuracy~\cite{snow}.

Our work is also related to previous techniques in class list generation and categorization~\cite{cascade,alloy}. These methods use crowdsourcing algorithms to generate taxonomies for textual data~\cite{cascade}, and a collaboration between humans and machine learning algorithms to cluster text~\cite{alloy}. Here, we want to group visually indistinguishable objects together regardless of their textual taxonomy. Thus, we cannot rely on textual data to create our class list of cars. Our task also differs from taxonomy creation in that category names are known beforehand. Thus, crowd workers only have to answer questions regarding objects' visual similarity. They are not required to create category names for objects. The objects we consider might also have very small visual distinctions making the categorization prone to errors by non-experts. Thus, we introduce a new crowdsourcing algorithm to create a visual taxonomy of synthetic fine grained categories.

\begin{figure}
 \centering
 \includegraphics[width=0.9\columnwidth]{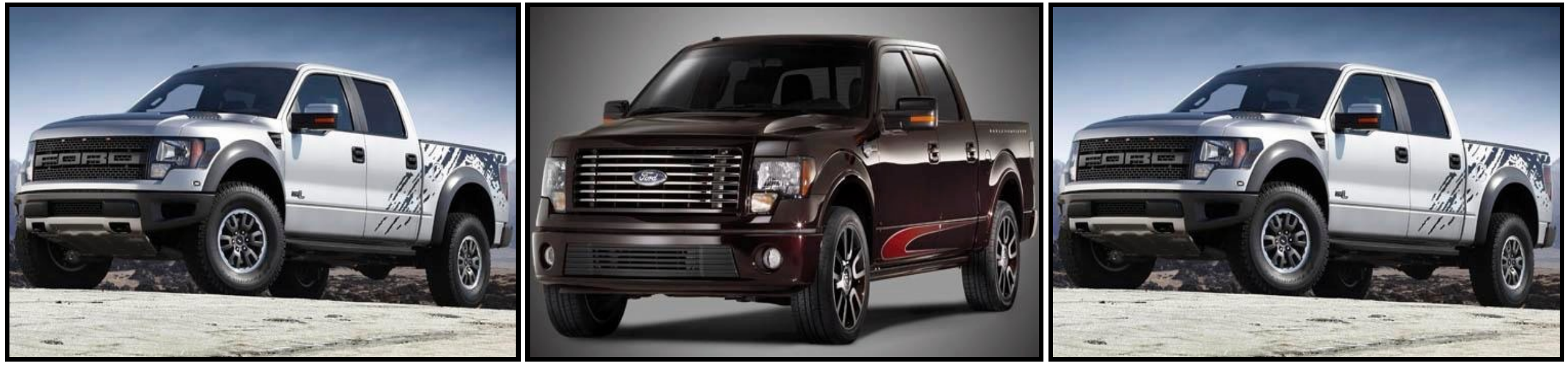}
 \caption{Unlike existing fine-grained image annotation methods, we cannot rely on taxonomy to group cars into visual categories. For instance, a 2011 Ford F-150 Supercrew HD (left) looks the same as a 2001 Ford F-150 Supercrew LL (right) but very different from a 2011 Ford F-150 Supercrew SVT (middle).}~\label{fig:figure2}
\end{figure}
\section{Approach}
\begin{figure*}[ht!]
 \centering
 \includegraphics[width=1.9\columnwidth]{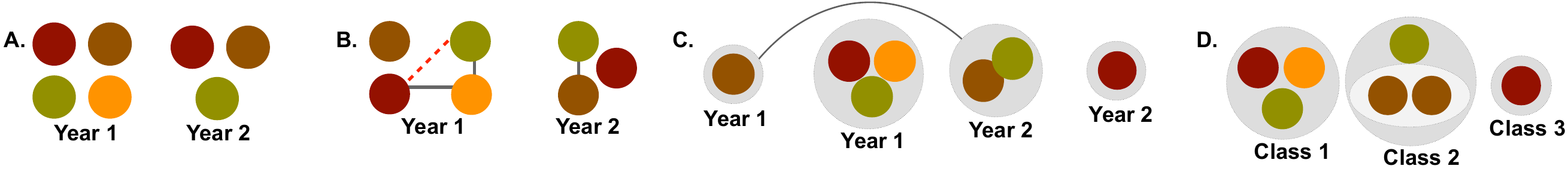}
 \caption{A simple example illustrating our procedure to cluster car categories. A. A graph with each circle representing a node. Different colors indicate distinct trims under the same make, model, body type and year of a car. We query whether all pairs of trims in Year 1 should be merged and do the same for Year 2. B. The resulting graph structure with an edge (in grey) connecting trims with the same appearance. A dashed line between red and green trims shows an edge that would have been present if the merging procedure for red, green and yellow trims in Year 1 were correct. After detecting that edges between red, green and yellow trims do not form a clique, we repeat the query for those nodes. C. The merging process for the same trims in different years. An edge is added between trims in Year 1 and Year 2 due to their indistinguishable visual appearance. D. The cars are grouped into 3 classes because there are 3 connected components in C.}~\label{fig:clustering}
\end{figure*}

\subsection{Constructing the class list} 
While current fine-grained datasets~\cite{cubs} rely on object taxonomy to create groupings of visually indistinguishable objects, we cannot use the same methodology to construct our class list of cars. For example, species of birds with the same name (like Bobolink) can be classified into one visual category whereas those with different names (like Bobolink and Cardinal) are visually different and belong in separate classes. However, different synthetic object categories are not necessarily correlated with their visual similarity. As seen in Figure~\ref{fig:figure2}, objects with different names can look the same (e.g.
 2011 Ford F-150 Supercrew HD and 2001 Ford F-150 Supercrew LL) 
 or different (2011 Ford F-150 Supercrew HD and 2001 Ford F-150 Supercrew SVT). We present a graph based crowdsourcing approach to creating this class list, distilling the crowdsourcing task to one simple binary question: ``Are these two cars the same?''. 

Cars follow a taxonomic hierarchy visualized in Figure~\ref{fig:car_hierarchy} where the root node of each tree in the forest is the make of the car (e.g. Honda vs. BMW). We know all nodes in one tree to be visually different from all nodes in a different tree. Under the make node is the model (e.g. 3 Series, 5 Series etc. for BMW). Under the model is the body type (e.g. sedan, wagon etc.) and manufactured year and trim (e.g. 330i, 335d, etc.). As noted earlier, nodes under one body type can look the same or different. Each type of car can be visualized as a leaf node in this tree. 

We group visually indistinguishable cars using a bottom-up clustering approach combined with a binary crowdsourcing task. Starting at the leaf nodes, we recursively query whether or not to combine pairs of child nodes, moving up the tree in a reverse breadth-first-search fashion. We represent each distinct car category as a node in a graph (V, E) with zero edges. In our implementation, we start with $15,213$ types of cars listed in edmunds.com corresponding to $15,213$ nodes in our graph. Each car category is accompanied by example images. To query whether two types of cars should be merged, we show example images and ask AMT workers if the two cars are the same. Each task has 6 such binary questions with 2 gold standards, inserted in random order for quality control. Every time a ``yes'' answer is returned from a task, we add an undirected edge between the two car nodes. Once all edges are queried, we use connected components~\cite{connected-components} to form clusters of nodes resulting in our final set of categories. 

\begin{figure}
\centering
\includegraphics[width=1\columnwidth]{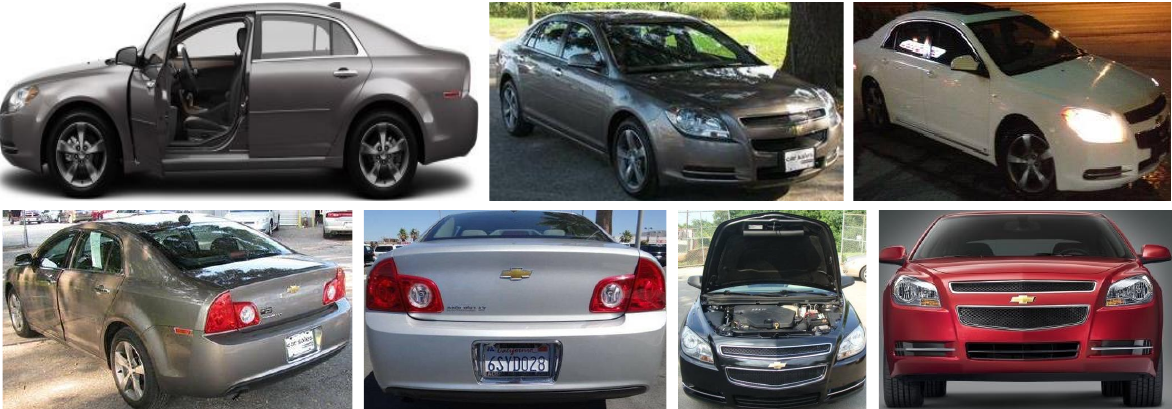}
\caption{Examples of images for 2008-2010 Chevrolet Malibu sedan ls,lt,base,fleet showing intra class variation in our dataset. In addition to variations in color, background, lighting and pose, doors and trunks can be open or closed and cars can be occluded.}
\label{fig:intraclass}
\end{figure}

Figure~\ref{fig:clustering} further describes the process with an example. We start with the graph in Figure~\ref{fig:clustering}A. where each node is a car with a specific make, model, year and trim (leaf nodes in the car taxonomy tree). We query whether trim 1 and trim 2 of the same make, model and year should be merged. If yes, we add an undirected edge between them (Figure~\ref{fig:clustering}B). We repeat this process comparing all trims within the same make, model and year adding edges as necessary. For quality control, we make use of the fact that all edges need to form cliques. That is, if car 1 and car 2 are determined to look the same, and car 2 and car 3 are also grouped together, then cars 1 and 3 should also be merged. We repeat the querying process for nodes that do not have this correct edge structure. 

Once we cluster all trims of cars in the same year, we arrive at the graph structure in Figure~\ref{fig:clustering}B. We repeat this process, comparing trims with the same name across different years and adding edges where necessary (Figure~\ref{fig:clustering}C). The output of this step is a graph where each node consists of merged car categories and each edge connects cars whose appearance has not changed throughout the compared years. We use connected components to get our final list of category groupings (Figure~\ref{fig:clustering}D), where the number of connected components corresponds to the number of classes. 

Although we used our methodology to collect annotated images for a car dataset, it is generalizable to other synthetic fine-grained domains with certain characteristics. Our workflow is effective because an exhaustive list of car classes is available to start the clustering process and there are image repositories (craigslist.com, cars.com) that can be queried with object names. Thus, our methodology is useful in other domains where such class lists and repositories are also available. These domains include most objects sold online like clothes (shoes, shirts, etc.), or other retail items such as those sold on Amazon (phones, etc.).

\subsection{Large Scale Fine-Grained Dataset} To automatically obtain car images with fine-grained labels, we leverage e-commerce websites like craigslist.com and cars.com. The images on these websites are uploaded by people hoping to sell their cars. Thus, they are likely to be labeled with the car make, model, body type, year and trim information allowing us to obtain labels for free. For each of the $15,213$ types of cars in our dataset, we construct a query by concatenating the make, model, body type, year and trim of the car. We enter each query into cars.com and craigslist.com and use regular expressions to analyze all returned posts. Finally, we download images associated with posts whose titles contain all the words in our query. We cleanup our dataset by crowdsourcing binary tasks, filtering out images that do not contain cars and using our implementation of~\cite{sheng} for quality control. We gathered a total number of $712,430$ annotated images for a total cost of only $\$5,000$.

Figure~\ref{fig:intraclass} shows example images highlighting our dataset's intraclass variability. All images have a single car that occupies most of the pixel space. In addition to variations in pose, lighting and resolution, some cars in the same class are occluded and others are deformed (due to accidents). They can also have open or closed doors, backs of trucks that are covered or uncovered, and additional ornaments. 

\section{Experiments}
We now estimate the accuracy of our constructed class list as well as the precision of our dataset. Then, we use it to train a fine-grained car classifier, publishing results that can serve as baselines for future researchers using our dataset.

To measure the accuracy of our constructed class list, we hired an expert to manually group all $92$ types of Honda Accord sedans in our dataset into visually indistinguishable sets of cars. Then, for each type of car in this set, we measured the intersection between vehicles placed in the same group as the car by our algorithm and by the expert. We divide the intersection by the union of other vehicles grouped with the car by both methods. 

Specifically, for each car type $c$, let $S_{exp}$ be the set of cars grouped into the same category as $c$ by the expert, and $S_{alg}$ be the set of cars grouped with $c$ by our method. We calculate 
\begin{equation}
(S_{exp} \cap S_{alg})/(S_{exp} \cup S_{alg})
\end{equation}
to measure our level of agreement with the expert in grouping each car. Averaging this number over the $92$ selected cars gives us an estimate of $81.09\%$ agreement with the expert across all categories in question.

To estimate the precision of our dataset, we look at $100$ random images for $170$ classes of cars, investigating a total of $17,000$ images. $96.6\%$ of the investigated images were of cars belonging to the right category verifying that our dataset has a high precision. To put this number into context: CUB 2011~\cite{cubs}, a standard fine-grained dataset today, has an accuracy of ~96\% and ImageNet's fine-grained categories have a higher annotation error~\cite{fine-print}. An in-depth analysis by~\cite{fine-print} shows that a 5\% annotation error minimally affects fine-grained classification accuracy when the number of categories approaches 1000. 

Finally, we trained a fine-grained car classification model following the method of~\cite{alexnet}, and tested it on all $2,657$ classes in our dataset. We obtained an accuracy of $67.60\%$ verifying that we have constructed a challenging fine-grained dataset with categories that are difficult to visually distinguish. The same method achieves a similar classification accuracy of 61.3\% on CUBS 2011~\cite{cubs}. This low baseline shows that our dataset will be useful in training and benchmarking algorithms that seek to improve fine-grained classification accuracy. Table~\ref{tab:results} summarizes our results.

\begin{table}
\begin{center}
\begin{tabular}{l l}
Experiment & Result\\
\hline\hline
Class list accuracy & 81.09\%\\
Dataset precision & 96.60\%\\
Fine-grained classification accuracy & 67.60\%\\
\end{tabular}
\end{center}
\caption{Experimental results measuring the accuracy of our constructed class list, the precision of our dataset and fine-grained car classification accuracy of a model trained with, and tested on our dataset.}~\label{tab:results}
\end{table}

\section{Conclusion}
We have presented a workflow to construct a dataset of visually similar synthetic objects -- in this case, cars -- without using experts. We used our workflow to construct a large-scale fine-grained dataset consisting of $712,430$ images of $2,657$ car categories comprising all cars manufactured since 1990 and annotated at a fraction of the cost ($\$6,000$) of hiring experts. Central to our approach is a graph based crowdsourcing algorithm to group different types of cars into visually distinct categories. This is an important and often difficult step because taxonomic divisions do not correspond to visual similarity in synthetic objects. What types of differences count must also be predefined by creators of the dataset, and communicated to crowd workers. Workers are asked to compare cars with no differences of interest, or differences that are very subtle. Thus, errors are mostly in recall rather than precision: if the worker indicates a difference, there usually is one. However, workers can miss subtle differences and indicate that $2$ objects are visually the same. We suggest future work investigating the use of our workflow to gather a richer dataset. For instance, instead of only answering whether two types of cars are visually the same, workers can indicate differences by drawing boxes on car parts.

\section{Acknowledgments}
We thank Ranjay Krishna whose continuous help and encouragement made this paper possible, and Olga Russakovsky and Dana\"{e} Metaxa-Kakavouli for their valuable feedback. This research is partially supported by an NSF grant (IIS-1115493), the Stanford DARE fellowship (to T.G.) and by NVIDIA (through donated GPUs).

\bibliographystyle{SIGCHI-Reference-Format}
\bibliography{sample}


\begin{thebibliography}{00}


\ifx \showCODEN    \undefined \def \showCODEN     #1{\unskip}     \fi
\ifx \showDOI      \undefined \def \showDOI       #1{{\tt DOI:}\penalty0{#1}\ }
  \fi
\ifx \showISBNx    \undefined \def \showISBNx     #1{\unskip}     \fi
\ifx \showISBNxiii \undefined \def \showISBNxiii  #1{\unskip}     \fi
\ifx \showISSN     \undefined \def \showISSN      #1{\unskip}     \fi
\ifx \showLCCN     \undefined \def \showLCCN      #1{\unskip}     \fi
\ifx \shownote     \undefined \def \shownote      #1{#1}          \fi
\ifx \showarticletitle \undefined \def \showarticletitle #1{#1}   \fi
\ifx \showURL      \undefined \def \showURL       #1{#1}          \fi

\bibitem{birdsnap}
{Thomas Berg}, {Jiongxin Liu}, {Seung~Woo Lee}, {Michelle~L Alexander},
  {David~W Jacobs}, {and} {Peter~N Belhumeur}. 2014.
\newblock \showarticletitle{Birdsnap: Large-scale fine-grained visual
  categorization of birds}. In {\em 2014 IEEE Conference on Computer Vision and
  Pattern Recognition}. IEEE, 2019--2026.
\newblock


\bibitem{fg4}
{Steve Branson}, {Grant Van~Horn}, {Serge Belongie}, {and} {Pietro Perona}.
  2014.
\newblock \showarticletitle{Bird species categorization using pose normalized
  deep convolutional nets}.
\newblock {\em arXiv preprint arXiv:1406.2952\/} (2014).
\newblock


\bibitem{alloy}
{Joseph~Chee Chang}, {Aniket Kittur}, {and} {Nathan Hahn}. 2016.
\newblock \showarticletitle{Alloy: Clustering with Crowds and Computation}. In
  {\em Proceedings of the 2016 CHI Conference on Human Factors in Computing
  Systems}. ACM, 3180--3191.
\newblock


\bibitem{cascade}
{Lydia~B Chilton}, {Greg Little}, {Darren Edge}, {Daniel~S Weld}, {and}
  {James~A Landay}. 2013.
\newblock \showarticletitle{Cascade: Crowdsourcing taxonomy creation}. In {\em
  Proceedings of the SIGCHI Conference on Human Factors in Computing Systems}.
  ACM, 1999--2008.
\newblock


\bibitem{imagenet}
{Jia Deng}, {Wei Dong}, {Richard Socher}, {Li-Jia Li}, {Kai Li}, {and} {Li
  Fei-Fei}. 2009.
\newblock \showarticletitle{Imagenet: A large-scale hierarchical image
  database}. In {\em Computer Vision and Pattern Recognition, 2009. CVPR 2009.
  IEEE Conference on}. IEEE, 248--255.
\newblock


\bibitem{jia}
{Jia Deng}, {Olga Russakovsky}, {Jonathan Krause}, {Michael~S Bernstein}, {Alex
  Berg}, {and} {Li Fei-Fei}. 2014.
\newblock \showarticletitle{Scalable multi-label annotation}. In {\em
  Proceedings of the SIGCHI Conference on Human Factors in Computing Systems}.
  ACM, 3099--3102.
\newblock


\bibitem{edmunds}
{Edmunds}. 2016.
\newblock edmunds.com.
\newblock   (2016).
\newblock
\newblock
\shownote{\url{http://www.edmunds.com/}.}


\bibitem{connected-components}
{John~E Hopcroft} {and} {Robert~E Tarjan}. 1971.
\newblock \showarticletitle{Efficient algorithms for graph manipulation}.
\newblock  (1971).
\newblock


\bibitem{fg1}
{Aditya Khosla}, {Nityananda Jayadevaprakash}, {Bangpeng Yao}, {and} {Li
  Fei-Fei}. 2011.
\newblock \showarticletitle{Novel dataset for fine-grained image
  categorization: Stanford dogs}. In {\em Proc. CVPR Workshop on Fine-Grained
  Visual Categorization (FGVC)}, Vol.~2.
\newblock


\bibitem{jon}
{Jonathan Krause}, {Benjamin Sapp}, {Andrew Howard}, {Howard Zhou}, {Alexander
  Toshev}, {Tom Duerig}, {James Philbin}, {and} {Li Fei-Fei}. 2015.
\newblock \showarticletitle{The Unreasonable Effectiveness of Noisy Data for
  Fine-Grained Recognition}.
\newblock {\em arXiv preprint arXiv:1511.06789\/} (2015).
\newblock


\bibitem{ranjay}
{Ranjay Krishna}, {Kenji Hata}, {Stephanie Chen}, {Joshua Kravitz}, {David~A
  Shamma}, {Li Fei-Fei}, {and} {Michael~S Bernstein}. 2016.
\newblock \showarticletitle{Embracing error to enable rapid crowdsourcing}.
\newblock {\em arXiv preprint arXiv:1602.04506\/} (2016).
\newblock


\bibitem{alexnet}
{Alex Krizhevsky}, {Ilya Sutskever}, {and} {Geoffrey~E Hinton}. 2012.
\newblock \showarticletitle{Imagenet classification with deep convolutional
  neural networks}. In {\em Advances in neural information processing systems}.
  1097--1105.
\newblock


\bibitem{fg2}
{Jiongxin Liu}, {Angjoo Kanazawa}, {David Jacobs}, {and} {Peter Belhumeur}.
  2012.
\newblock \showarticletitle{Dog breed classification using part localization}.
  In {\em European Conference on Computer Vision}. Springer, 172--185.
\newblock


\bibitem{visipedia}
{Pietro Perona}. 2010.
\newblock \showarticletitle{Vision of a Visipedia}.
\newblock {\it Proc. IEEE} {98}, 8 (2010), 1526--1534.
\newblock


\bibitem{sheng}
{Victor~S Sheng}, {Foster Provost}, {and} {Panagiotis~G Ipeirotis}. 2008.
\newblock \showarticletitle{Get another label? improving data quality and data
  mining using multiple, noisy labelers}. In {\em Proceedings of the 14th ACM
  SIGKDD international conference on Knowledge discovery and data mining}. ACM,
  614--622.
\newblock


\bibitem{vgg}
{Karen Simonyan} {and} {Andrew Zisserman}. 2014.
\newblock \showarticletitle{Very deep convolutional networks for large-scale
  image recognition}.
\newblock {\em arXiv preprint arXiv:1409.1556\/} (2014).
\newblock


\bibitem{snow}
{Rion Snow}, {Brendan O'Connor}, {Daniel Jurafsky}, {and} {Andrew~Y Ng}. 2008.
\newblock \showarticletitle{Cheap and fast---but is it good?: evaluating
  non-expert annotations for natural language tasks}. In {\em Proceedings of
  the conference on empirical methods in natural language processing}.
  Association for Computational Linguistics, 254--263.
\newblock


\bibitem{fine-print}
{Grant Van~Horn}, {Steve Branson}, {Ryan Farrell}, {Scott Haber}, {Jessie
  Barry}, {Panos Ipeirotis}, {Pietro Perona}, {and} {Serge Belongie}. 2015.
\newblock \showarticletitle{Building a bird recognition app and large scale
  dataset with citizen scientists: The fine print in fine-grained dataset
  collection}. In {\em Proceedings of the IEEE Conference on Computer Vision
  and Pattern Recognition}. 595--604.
\newblock


\bibitem{fg5}
{Catherine Wah}, {Steve Branson}, {Peter Welinder}, {Pietro Perona}, {and}
  {Serge Belongie}. 2011.
\newblock \showarticletitle{The caltech-ucsd birds-200-2011 dataset}.
\newblock  (2011).
\newblock


\bibitem{cubs}
{P. Welinder}, {S. Branson}, {T. Mita}, {C. Wah}, {F. Schroff}, {S. Belongie},
  {and} {P. Perona}. 2010a.
\newblock {\em {Caltech-UCSD Birds 200}}.
\newblock {T}echnical {R}eport CNS-TR-2010-001. California Institute of
  Technology.
\newblock


\bibitem{welinder}
{Peter Welinder}, {Steve Branson}, {Pietro Perona}, {and} {Serge~J Belongie}.
  2010b.
\newblock \showarticletitle{The multidimensional wisdom of crowds}. In {\em
  Advances in neural information processing systems}. 2424--2432.
\newblock


\bibitem{compcars}
{Linjie Yang}, {Ping Luo}, {Chen Change~Loy}, {and} {Xiaoou Tang}. 2015.
\newblock \showarticletitle{A large-scale car dataset for fine-grained
  categorization and verification}. In {\em Proceedings of the IEEE Conference
  on Computer Vision and Pattern Recognition}. 3973--3981.
\newblock


\end{thebibliography}

\end{document}